\documentclass[aps,amsmath,amssymb,floatfix,superscriptaddress,nofootinbib,12pt]{article}
\pdfoutput=1 % if your are submitting a pdflatex (i.e. if you have
\usepackage{jheppub} % for details on the use of the package, please
\usepackage{youngtab}
\usepackage{hyperref}
\usepackage{fancybox}
\usepackage{float}
\usepackage{amsfonts}
\usepackage{amsmath}
\usepackage{amsbsy}
\usepackage{graphicx}
\usepackage{amssymb}
\usepackage{amsthm}
\usepackage{bm}
\usepackage{comment}
\usepackage{epsfig}
\usepackage{latexsym}
\usepackage{pdflscape}
\usepackage{color}
\usepackage{mathtools}
\makeatletter
\def\@fpheader{\relax}
\makeatother

\DeclareSymbolFont{AMSa}{U}{msa}{m}{n}
\DeclareSymbolFont{AMSb}{U}{msb}{m}{n}
\DeclareMathSymbol{\fieldR}{\mathalpha}{AMSb}{"52}

\def\hri#1#2{\href{http://arxiv.org/abs/#1}{[ArXiv:#1]#2}}

\def\hrj#1#2{\href{https://doi.org/#1}{#2}}

\DeclareMathOperator{\tr}{tr}

\newcommand{\beq}{\begin{eqnarray}}
\newcommand{\eeq}{\end{eqnarray}}
\newcommand{\bea}{\begin{eqnarray}}
\newcommand{\eea}{\end{eqnarray}}
\newcommand{\be}{\begin{equation}}
\newcommand{\ee}{\end{equation}}
\newcommand{\bq}{\begin{equation}}
\newcommand{\eq}{\end{equation}}

\let\svthefootnote\thefootnote
\newcommand\blankfootnote[1]{%
  \let\thefootnote\relax\footnotetext{#1}%
  \let\thefootnote\svthefootnote%
}
\def\6{\partial}

\def\a{\alpha}

\def\tr{{\rm Tr}}

\def\6{\partial}

\title{Baby Universes born from the Void}
\author{Panos Betzios$^{1}$, }
\author{Nava Gaddam$^{2}$, and} 
\author{Olga Papadoulaki$^3$}

\affiliation[1]{\href{https://phas.ubc.ca/}{Department of Physics and Astronomy}, University of British Columbia,
6224 Agricultural Road, Vancouver, B.C. V6T 1Z1, Canada.}

\affiliation[2]{\href{https://www.uu.nl/en/research/institute-for-theoretical-physics}{Institute for Theoretical Physics} and Center for Extreme Matter and Emergent Phenomena, Utrecht University, 3508 TD Utrecht, The Netherlands.}

\affiliation[3]{\href{https://perimeterinstitute.ca/}{Perimeter Institute for Theoretical Physics}, Waterloo, Ontario N2L 2Y5, Canada. \vspace{0.5cm}}

\emailAdd{pbetzios@phas.ubc.ca}   
\emailAdd{gaddam@uu.nl}
\emailAdd{opapadoulaki@perimeterinstitute.ca \newline}

\abstract{We propose a novel construction of a third quantised baby universe Hilbert space $\mathcal{H}_{BU}$ for the quantum gravity path integral. In contrast to the original description of $\a$-parameters, both the bulk and boundary microscopic parameters are fixed in our proposal. Wormholes and baby universes appear due to refined observables, of the boundary dual quantum field theories, that crucially involve the space of representations of the boundary gauge group. Irreducible representations, on which the path integral factorises, give rise to field theoretic superselection sectors and replace the $\a$ states. } 
\begin{document} 
\maketitle
\flushbottom

%\clearpage

\setcounter{page}{1}

\section{Introduction}

Among the most important open problems in theoretical physics, the role of topology change in quantum gravity is particularly tantalising. The intrigue has only grown with recent interest \cite{Bousso:2022ntt} in spacetime wormholes connecting multiple asymptotic boundaries. In the late eighties, it was argued that the effects of wormholes are accessible in the effective low energy theory via averaged couplings to local operators \cite{Coleman:1988cy,Lavrelashvili:1987jg,Hawking:1988ae,Giddings:1988cx}; these couplings were called the $\alpha$-parameters. The unrealised initial hope was that the integral over the $\alpha$-parameters would be sharply peaked in favour of an infinitesimal value for the observed positive cosmological constant. Nevertheless, these ideas led to a novel \emph{third quantised picture} of ``baby universes (BU)'' that are created and destroyed in superspace \cite{Giddings:1988wv}, resulting in dynamical topology changing processes that affect the physics of a large semi-classical ``parent manifold''\textemdash the one we observe at large scales.

On the other hand, it has long been expected that there are no free parameters in a quantum theory of gravity and that all symmetries are gauged \cite{Krauss:1988zc, Kallosh:1995hi, Banks:2010zn}. This expectation was further strengthened by the advent of holography \cite{Maldacena:1997re, Witten:1998qj}, where in all controllable higher dimensional\footnote{In this essay, by ``higher dimensions'' we mean $d>2$.} examples of string/M-theory, all the parameters on which the boundary field theory depends are known and fixed. Should the holographic duality hold, the bulk dual quantum gravity (QG) theory must also depend on the same fixed quantities, leaving no room for $\alpha$-parameters. Several authors have since emphasised the apparent conflict between holography and the existence of $\alpha$-parameters \cite{Rey:1998yx, ArkaniHamed:2007js,Hebecker:2018ofv}, and between holography and fixed multiboundary saddles \cite{Maldacena:2004rf, Hertog:2018kbz}. On the contrary, others have suggested a harmony in simple topological \cite{Marolf:2020xie} or $2d$ models \cite{Blommaert:2022ucs,Johnson:2022wsr}. However, known UV complete variants of these models imply that their non-trivial topological features arise on the string worldsheet rather than the target space \cite{Betzios:2020nry, McNamara:2020uza}. Whereas in higher dimensions, Euclidean wormholes also arise in bulk duals of appropriately \emph{interacting} QFT's \cite{worm, VanRaamsdonk:2020tlr, Betzios:2021fnm}, but these cases leave no room for conventional $\a$-parameters.

In this essay, we provide a bridge between these two apparently conflicting descriptions of the QG path integral in higher dimensions. Before we do so, we will first describe why the conventional picture of $\a$-parameters fails, necessitating an alternative approach.

\section{The failure of the bilocal description and $\a$-parameters}\label{sec:bilocal_failure}

Coleman's $\a$-states emerge from a bi-local expansion of the contribution of Euclidean wormholes to the gravitational path integral. This expansion is presumed to emerge in the limit of soft metric perturbations (whose wavelengths far exceed the size of the ends of the wormholes (instantons)). In such a limit, the two instanton ends of a wormhole were argued to couple to local bulk operators while preserving their connection to each other, resulting in a bilocal action. This bilocal action can then be traded for averaged $\alpha$-parameters. This perspective has been reviewed in \cite{Hebecker:2018ofv, Kundu:2021nwp}, for instance. Let us now analyse the assumptions going into this emergence of $\a$-parameters. To this end, let us define the following length scales\footnote{We assume Planck length $L_{P}$ to be the smallest of all scales in the theory, for validity of effective field theory.}: 
\begin{itemize}
\item $L_{inst}$ defines the size of the instanton (or the ends of the wormhole).
\item $L_{th}$ is the shortest scale associated with the wormhole (it dictates the size of the wormhole throat).
\item $L_w$ is the length of the wormhole (which defines the distance between the two ends of the wormhole).
\item And finally, $L_{probe}$ is the scale at which we probe the theory.
\end{itemize}
In all known solutions, there is a natural hierarchy among these scales $L_P \ll L_{th} < L_{inst} < L_w$. One may further argue more generally that even certain off shell configurations (such as the double cone of \cite{Saad:2019lba}) satisfy $L_{th} < L_{inst}$, since the throat is always the minimal characteristic scale of the wormhole. The existence of a bilocal action, and consequently the emergence of averaged $\a$ parameters, relies on the following assumptions:

\begin{enumerate}

\item A local operator expansion that effectively replaces each wormhole-end or instanton with an infinite sum of local bulk operators $\sum_i c_i \int d^{d+1} x \sqrt{g(x)}  \mathcal{O}_i(x)$. This is achieved in the long wavelength limit when the theory is probed at scales much larger than the size of the instanton $L_{probe} \gg L_{inst}$. This implies that the coefficients $c_{i}$ must be defined in powers of ${L_{inst}}/L_{probe}$.

\item Validity of the dilute gas approximation. This means that $L_{inst} \ll L_w$, so that the two endpoint instantons never overlap.

\item The throat must retain a connection between the two instantons or wormhole endpoints. Moreover, this connection must not depend on the location of the endpoints. Then, the effective action takes the bi-local form
\be
S_{EFT} ~ = ~ \int d^{d+1} x \sqrt{g(x)} \, \int d^{d+1} y \sqrt{g(y)} \,  \sum_{i, j } \, \Delta_{i j}  \, \mathcal{O}_i(x) \, \mathcal{O}_j(y) \,.
\ee
The path integral containing this bilocal action can be recast in terms of auxiliary $\alpha$-parameters.
\end{enumerate}
We now immediately observe that these conditions cannot all be simultaneously satisfied. Indeed, since the size of the throat is always necessarily smaller than the size of the instanton, $L_{th} < L_{inst}$, we see that the throat pinches off before the instantons can be replaced by local operators. Therefore, in this limit, the wormhole path integral factorises, resulting in two \emph{disconnected} local operator expansions at each endpoint. This does not mean that wormholes cannot exist, it merely implies that they will not be visible in the infrared where the endpoints are treated as local operator insertions.

Therefore, we conclude that the bilocal expansion has zero radius of validity in parameter space, leaving no room for conventional $\a$-parameters even from the point of an IR expansion. Of course, all notions of geometry break down in the UV.

\section{Our proposal and motivations}

Given the failure of the bilocal expansion described in the previous section, we now make the following proposal for the Hilbert space of baby universes.

\subsection{Proposal} In addition to the familiar Hilbert space of asymptotically AdS states in holography, the QG path integral also describes observables that belong in a \emph{non trivial} third quantised baby universe Hilbert space ($\mathcal{H}_{BU}$). The latter gives rise to effective superselection sectors through the decomposition $\mathcal{H}_{BU} = \oplus_R \, \mathcal{H}_R$ in terms of unitary representations, $R$, of the compact (gauge) group symmetry $\mathcal{G}$. This refinement is incorporated into the dual field theory partition function via the insertion of appropriate non-local operators that transform in the said representations $R$. We thereby arrive at an augmentation of the space of field theoretic observables which is important for the complete description of the bulk gravitational physics.

\subsection{Motivations for our proposal} 

We will make use of several known examples to both motivate and elucidate various aspects of our proposal.

\paragraph{Liouville theory} The state-operator correspondence is subtle in Liouville quantum gravity owing to the presence of both microscopic and macroscopic bulk states \cite{Seiberg:1990eb}. In analogy\footnote{\label{fnote:tHooft} That the QG path integral should be viewed as a higher dimensional version of Liouville theory with a dynamical conformal factor was proposed in \cite{tHooft:2014swy}.}, we define macroscopic operators\footnote{In matrix models, these are the familiar loop operators \cite{Saad:2019lba,Betzios:2020nry,McNamara:2020uza}.} $\hat{Z}[J]$ as functionals of sources, which create large $AdS$ conformal boundaries. These are associated with partition functions (Schwinger functionals) of the holographic dual field theory. In contrast, we introduce microscopic ``holes'' or ``vacuoles'' in the geometry\footnote{In the $L_{Pl} \rightarrow 0$ limit, these can be thought of as ``defects" or ``punctures" of the geometry. A similar idea was explored in \cite{Betzios:2017krj}.} that correspond to microscopic states. A basis for describing these is provided by the representation eigenstates $|R\rangle$ of the gauge group; we propose that this replaces the $\alpha$-state basis. That is to say that the bulk state-operator correspondence in any UV complete theory of quantum gravity bears a resemblance to Liouville theory and not to non-gravitating CFT's or topological theories analysed in \cite{Marolf:2020xie}.

\paragraph{Holography} The cogent success of gauge/gravity duality leaves little room for averaged $\alpha$-parameters that are left unfixed. This motivates our proposal to abandon the $\alpha$ parameters in favour of microscopic states labelled by the representations $|R\rangle$, which do \emph{not} couple to local operators in the bulk effective action. In contrast to the proposal of \cite{Marolf:2020xie, McNamara:2020uza}, where the Hartle-Hawking state was expanded in a basis of $\alpha$-eigenstates, our proposal is to expand the dynamical
QG vacuum in a basis of $|R\rangle$ with the Hartle-Hawking state being the trivial or singlet representation. Therefore, there is a natural selection rule for our proposed macroscopic $\hat{Z}[J]$ operators that separates different representation eigenstates. This implies a superselection rule for all observables in the Hilbert space $\mathcal{H}_{R}$ associated with each macroscopic boundary arising from the expectation value of $\hat{Z}[J]$ in the state $|R\rangle$. We describe this further in Section \ref{sec:techdetails}.

\paragraph{Non-singlet sectors} Our proposal is a significant departure from the original description of $\a$-parameters as random multipliers of local bulk operators that result in randomness in the parameters of the dual field theory. In our proposal, all the microscopic parameters are fixed. Moreover, since the boundary field theory group is a gauge group, only gauge invariant states  (``singlets") must appear in the partition function. Therefore, the non-trivial representations are accompanied by a insertions of appropriate operators (characters) in the field theory path integral\footnote{Analyses of non-singlet states in holography have been reviewed by Callan in \cite{Mtheory}.}. We describe further technical details in Section \ref{sec:techdetails}.

\paragraph{Energy scale of topology change} Topology changing processes in quantum gravity are often expected to be relevant near Planck scale. However, it was shown in \cite{Tong:2014era} that they can in fact emerge at exponentially suppressed scales allowing them to be accessed in the effective bulk theory. Moreover, the non-trivial topologies were shown to renormalise the bare coupling constant of concern. This lends further credibility to the proposal that coupling constants are fixed, and not averaged. Furthermore, it motivates our proposal for the existence of non-trivial macroscopic saddles  in the sum over representations. Potential perturbative stability of Euclidean wormholes \cite{Marolf:2021kjc, Loges:2022nuw} provides further encouragement for the contribution of such geometries to the quantum gravity path integral.

\paragraph{Topological strings and LLM geometries} Observables such as Wilson loops backreact on the geometry when they are in large representations (when the corresponding Young-Tableaux contain $O(N^2)$ boxes)~\cite{Lunin:2006xr}. For every such representation, the bulk dual has a different topology. A sum over all representations in the partition function would correspond to a sum over the various topologies, giving rise to a picture of ``spacetime foam". This can be made more precise in the context of topological strings~\cite{Iqbal:2003ds,Dijkgraaf:2005bp} and LLM geometries~\cite{Lin:2004nb,Lunin:2006xr}.

\begin{figure}[h!]
\begin{center}
\includegraphics[width=0.9\textwidth]{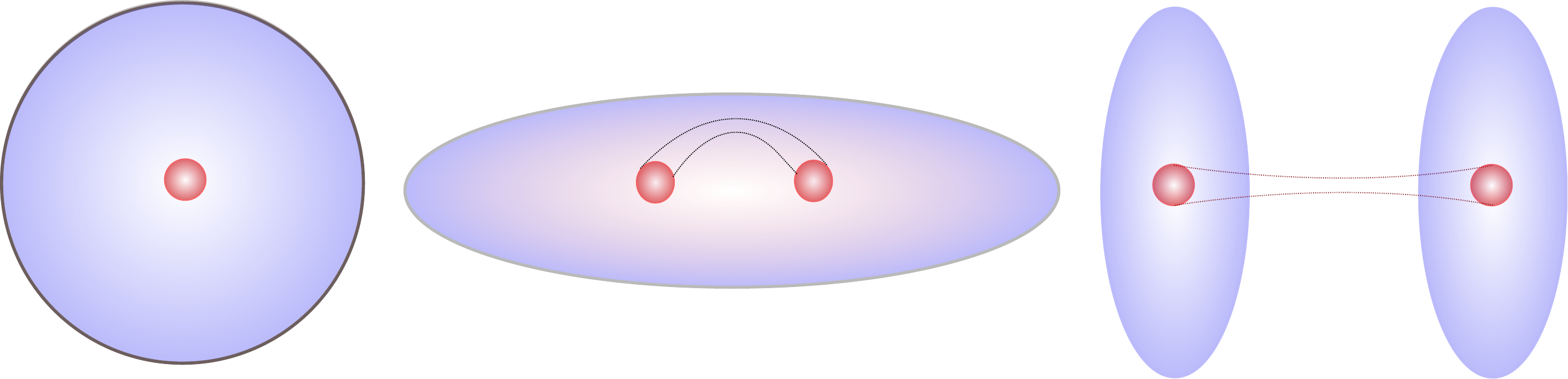}
\end{center}
\caption{In the far left, we depict the excision of a microscopic ``vacuole'' at the centre of global EAdS (whose topology is that of a ball). On the far right, two large macroscopic universes are connected by a microscopic wormhole whereas such a connection between distant locations in the same macroscopic universe is shown in the middle. Factorisation is restored in the \emph{representation} basis $| R \rangle$ (which we propose to replace the $\alpha$-basis prescription) for the BU Hilbert space $\mathcal{H}_{BU}$.}
\label{fig:wormvacuoles}
\end{figure}

We conclude this section with a pictorial illustration of our proposal in Fig. \ref{fig:wormvacuoles}. The conformal compactification of Euclidean $AdS_{d+1}$ is a $d+1$ dimensional ball with one macroscopic conformal boundary. Consider excising a small ball in the center, creating a ``vacuole" which is a microscopic (and not conformal) boundary. It is unnatural to relate this vacuole to a state that belongs in the usual field theoretic Hilbert space on the macroscopic conformal boundary, as the bulk QG theory is neither topological nor conformal\footnote{It may, however, be ``conformal" in a generalised Liouville sense. See Footnote \ref{fnote:tHooft}.}. Instead, we propose that such microscopic boundaries are associated with states in $\mathcal{H}_{BU}$, which is spanned by group representation eigenstates\footnote{The representation basis is \emph{not} a basis with a fixed number of microscopic boundaries. Instead, it replaces the $\a$-state basis in spirit. It would be interesting to understand if there exists a basis that describes a fixed number of vacuoles.}. From this perspective, $\hat{Z}[J]$ are operators that create macroscopic boundaries on the ``parent" manifold and do not belong to the microscopic third quantised BU Hilbert space $\mathcal{H}_{BU}$.

\section{Some technical details}\label{sec:techdetails}

We define the ``void" state in $\mathcal{H}_{BU}$ - the analogue of the vacuum \cite{Giddings:1988wv} - , and denote it by $| \Omega \rangle$. This state can be expanded in the basis of representation eigenstates as\footnote{The overlaps $\langle R | \Omega \rangle$ could also involve other objects such as Casimirs. Here we made the minimal choice that simply involves the degeneracy of the representations. }
\be
| \Omega \rangle = \sum_R \, D_R \, | R \rangle \quad \text{with} \quad \langle R | \Omega \rangle = D_R \, ,
\ee
where $D_R$ counts the degenerate states in each irreducible representation (the dimension). Given a concrete model, the ``void" may be constrained to all allowed representations with a finite number of boxes in the Young Tableaux\footnote{This is a natural way to restrict the space of representations for $U(N)$ gauge groups which come with a maximum number of rows in the Young diagrams. Discrete gauge groups have a naturally finite dimensional space of irreducible representations.}. In this case, the dimension of $\mathcal{H}_{BU}$ would be finite dimensional. In either case, for compact gauge groups, the Peter-Weyl theorem guarantees that this basis spans the total Hilbert space. In what follows, we divide all amplitudes by the norm of the void state\footnote{By our definition this counts the entropy of the collection of microscopic closed baby universes. The Hartle-Hawking state $|HH \rangle$ for a single microscopic boundary carries no entropy as in~\cite{McNamara:2020uza} and corresponds to the trivial or singlet representation.}. 

As a first non-trivial example, a single macroscopic boundary can now be expressed as an expectation value of a macroscopic operator
\be\label{HHexp}
Z[J]  \coloneqq   \frac{\langle \Omega | \hat{Z}[J] | \Omega \rangle}{\langle \Omega | \Omega \rangle}   =  \sum_R p_R Z_R[J] \quad \text{where} \quad p_R \coloneqq \frac{D_R^2}{\sum_{R'} D_{R'}^2} \, ,
\ee
and $Z_R[J] \coloneqq  \langle R | \hat{Z}[J] | R \rangle$. The spirit of the description of the $\alpha$-state path integral is now evident, whilst retaining fixed parameters for the boundary theory; $Z[J]$ implicitly depends on these fixed parameters. The familiar holographic source functional is merely the singlet term in \eqref{HHexp}. Since the boundary dual must necessarily only describe gauge invariant states, the astute reader might object to a definition of a quantity such as $Z_R[J]$. A purely field theoretic definition of such a quantity requires a modification of the boundary field theory functional, with the insertion of an appropriate (non-local) operator that depends on both the boundary gauge field (say $A^{a}_{\mu}$) and the representation $R$. For instance, such an operator would be the Wilson loop, or the character\footnote{Of course, the character depends on the holonomy of the gauge field around the circle (zero mode) and this can be integrated over in the path integral i.e. it is not a source $J$.} $\chi_R(U) \coloneqq \tr_{R} U = \tr_R e^{i \oint A}$, in gauged matrix quantum mechanics on a circle \cite{Betzios:2017yms, Betzios:2021fnm}. The integral over the gauge field zero mode projects the path integral onto the representation $R$. A two-dimensional generalisation for theories on a torus involves the Weyl-Kac characters \cite{Betzios:2021fnm}. Therefore, according to our proposal, the existence of bulk Euclidean wormholes is intimately tied to the existence and construction of such non-local operators. Provided a construction, it is then possible to define the source functional $Z_R[J]$, transforming in the representation $R$. Consequently, the information contained in $\mathcal{H}_{BU}$ is encapsulated in general (non-local) observables of the dual field theory; they cannot be accessed by the insertion of sources coupled to local gauge invariant operators in the Schwinger functional.

It is now straightforward to extend our construction to describe geometries with multiple conformal boundaries by inserting additional $\hat{Z}[J]$ operators. The two boundary case of Fig. \ref{fig:wormvacuoles}, for instance, corresponds to a state contributing to the expectation value
\be\label{twoboundary}
\langle \Omega | \hat{Z}^{(1)}[J_1] \, \hat{Z}^{(2)}[J_2] | \Omega \rangle ~ = ~ \sum_R \, p_R \, Z^{(1)}_R [J_1] \,  Z^{(2)}_R [J_2] \, .
\ee
The natural question now is to ask whether this equation contains geometric Euclidean wormhole(s) connecting the two large macroscopic universes. This question is equivalent to asking if the sum over group representations in \eqref{twoboundary} localises on a saddle with the effective characteristics of a macroscopic geometry connecting the two asymptotic AdS regions. The answer to this dynamical question is model dependent\footnote{See \cite{Betzios:2021fnm} for an analysis in concrete lower dimensional models.}. Nevertheless, some general comments are in order. Since geometric characteristics of such saddles are only expected to emerge at large $N$, small representations (with only a few number of Young-Tableau boxes) do not contribute; they may at best give rise to non-geometric microscopic quantum wormholes. A saddle point analysis therefore requires an appropriate limit in the space of representations (continuous Young-Tabeau diagrams) and study of the resulting equations \cite{Douglas:1993iia}.

\section{Consequences of our proposal}

\subsection{Comparisons with McNamara-Vafa and Marolf-Maxfield}

Motivated by swampland constraints, McNamara and Vafa formulated a ``baby universe hypothesis" \cite{McNamara:2019rup,McNamara:2020uza} which says that $\text{dim} \mathcal{H}_{BU} = 1$ in $d>2$ dimensions. This is further strengthened by the observation that a microscopic closed universe must have no entropy\footnote{ In our proposal the state of a single microscopic universe $|HH \rangle$ is described by the unique singlet representation and does not correspond to the BU ``void" $| \Omega \rangle$ .}. Then, there can be no superselection sectors at finite volume leaving no room for randomised $\alpha$-parameters which result in global symmetries in QG. Indeed, as we argued in Section \ref{sec:bilocal_failure}, if one demands a bilocal expansion in the effective field theory, the wormhole throat pinches off, leaving only the no-boundary state behind. In contrast, it was later shown in \cite{Betzios:2021fnm} that certain superselection sectors could be obtained in concrete unitary field theoretic holographic models at finite volume without averaging over couplings. These superselection sectors are in correspondence with the representations of the gauge group of the dual holographic field theory and replace the usual $\a$-states, as we propose in this essay. This inspires us to formulate the following extension: 
\paragraph{Extended baby universe hypothesis} ``The constraints of gravity are so strong that the baby universe Hilbert space $\mathcal{H}_{BU}$ is finite dimensional (but non-trivial). The $\alpha$-states dictating the superselection sectors are replaced by representation eigenstates $| R \rangle$ of the compact gauge group of the dual field theory. The number of (degenerate) states in any such superselection sector is counted by the dimension of the representation."

In \cite{Marolf:2020xie}, Marolf and Maxfield proposed that the baby universe Hilbert space is spanned by boundary creating operators. In contrast, our proposal distinguishes between microscopic and macroscopic operators. The microscopic operators span the baby universe Hilbert space. Another basis for this Hilbert space is the representation eigenstate basis (which replaces the $\alpha$-state basis in our proposal), in which expectation values of macroscopic operators factorise. The macroscopic operators create macroscopic $AdS$ boundaries. Therefore, our proposal implies that the state-operator correspondence is more subtle in gravitating theories than in the topological models considered in their work.

\subsection{Global symmetries and the Weak gravity conjecture}

Coleman observed \cite{Coleman:1988cy} that in the presence of random $\alpha$-parameters, baby universes carry charges associated with global symmetries resulting in superselection sectors with different values for the said charges. As we described above, our proposal replaces the familiar $\alpha$-states with representation eigenstates. Superselection sectors arise in this representation eigenbasis, with fixed representations. Therefore, there is no room for randomised undetermined parameters in our proposal. This ensures consistency with the expectation that there are no global symmetries in the bulk QG theory\footnote{Boundary global symmetries are of course allowed, which are accompanied by propagating gauge fields in the bulk.}. Moreover, it allows us to formulate a \emph{Generalised Completeness Hypothesis}, which states that in addition to the familiar Completeness Hypothesis \cite{Banks:2010zn, Casini:2021zgr}, every allowed representation is present in the spectrum of the Baby Universe Hilbert Space (see also \cite{McNamara:2021cuo}). Evidently, this is a corollary of the Peter-Weyl theorem. It is therefore tempting to argue that models with $\alpha$-parameters of Coleman belong in the swampland.

\subsection{A top-down perspective from $D$-branes}

Our proposal is reminiscent of constructions in string theory, arising from stacks of $D$-branes. Consider well-separated stacks of many backreacting $D$-branes. Each stack results in an $AdS$ geometry of its own. The singlet sector of the theory is described by factorised boundary theories in the strict decoupling limit. However, massive string excitations extending between the stacks are captured by the non-singlet sectors which in the low energy are described by $T$-branes \cite{Cecotti:2010bp}. It would be interesting to use constructions of intersecting branes \cite{Gaiotto:2008ak, VanRaamsdonk:2020tlr, Heckman:2021vzx}, together with the use of appropriate operators such as monopole operators, to explore our proposed representation eigenbasis in detail.

\subsection{Evading the Fischler-Susskind-Kaplunovsky catastrophe}

In \cite{Fischler:1988ia}, an IR catastrophe was anticipated owing to the existence of an overdensity of giant wormholes of macroscopic sizes. In our construction, there is no room for free moduli governing the size of the wormhole ``instantons" or throats (their sizes are fixed in terms of the microscopic parameters defining the model). Moreover, the density of wormholes is never expected to become large due to a version of the stringy exclusion principle\textemdash the size of the large parent manifold, and consequently that of the representations, depends on the rank of the gauge group. While a detailed analysis is necessary to demonstrate how the catastrophe is evaded explicitly, the premises leading to it are evidently not satisfied.

\subsection{Geometry and entanglement (of representations)}

While it is well established that a geometric connection in Lorentzian signature can be associated to the presence of non-trivial entanglement, such as in the case of the thermofield double state~\cite{Maldacena:2001kr,VanRaamsdonk:2010pw} dual to the eternal two-sided black hole, it does not suffice to explain the existence of Euclidean wormholes. A simple demonstration of this fact involves the analytic continuation of the Kruskal manifold which turns the Einstein-Rosen bridge into two disconnected Euclidean cigar geometries~\cite{worm}. Our proposal suggests a form of \emph{representation theoretic entanglement} in the purely Euclidean context~\cite{Betzios:2021fnm}. A simple demonstration of this is provided by the two boundary case of eqn. \eqref{twoboundary}, where the allowed field theoretic states belong in the ``entangled sum" $\sum_{R}\mathcal{H}_R^1 \otimes \mathcal{H}_R^2$.

\section*{Acknowledgements}

We thank Elias Kiritsis, Mark van Raamsdonk, and Gerard 't Hooft for various discussions related to the concerns of this essay over the years. We would also like to acknowledge helpful conversations with Thomas Grimm, Ji Hoon Lee, Stefano Lanza, Javi Magan, Juan Maldacena, Jake McNamara, Miguel Montero, and the string theory groups at Perimeter Institute, the University of British Columbia, and Utrecht University.

\noindent The research of P.B. is supported in part by the Natural Sciences and Engineering Research Council of Canada. P.B. and O.P. acknowledge support by the Simons foundation. Research at Perimeter Institute is supported in part by the Government of Canada through the Department of Innovation, Science and Economic Development and by the Province of Ontario through the Ministry of Colleges and Universities.
The research of NG is supported by the Delta-Institute for Theoretical Physics (D-ITP) that is funded by the Dutch Ministry of Education, Culture and Science (OCW).


\begin{thebibliography}{99}




\bibitem{Bousso:2022ntt}
R.~Bousso, X.~Dong, N.~Engelhardt, T.~Faulkner, T.~Hartman, S.~H.~Shenker and D.~Stanford,
{\em ``Snowmass White Paper: Quantum Aspects of Black Holes and the Emergence of Spacetime,''}
\hri{2201.03096}{ [hep-th]}.



 %\cite{Coleman:1988cy}
\bibitem{Coleman:1988cy}
  S.~R.~Coleman,
  {\em ``Black Holes as Red Herrings: Topological Fluctuations and the Loss of Quantum Coherence,''}
  \href{https://www.sciencedirect.com/science/article/pii/0550321388901101}{Nucl.\ Phys.\ B {\bf 307} (1988) 867.}
  %doi:10.1016/0550-3213(88)90110-1
  %%CITATION = doi:10.1016/0550-3213(88)90110-1;%%
  %406 citations counted in INSPIRE as of 29 Dec 2018

%\cite{Lavrelashvili:1987jg}
\bibitem{Lavrelashvili:1987jg}
  G.~V.~Lavrelashvili, V.~A.~Rubakov and P.~G.~Tinyakov,
  {\em ``Disruption of Quantum Coherence upon a Change in Spatial Topology in Quantum Gravity,''}
  \href{http://www.jetpletters.ac.ru/ps/index-v-46_en.shtml}{JETP Lett.\  {\bf 46} (1987) 167
   [Pisma Zh.\ Eksp.\ Teor.\ Fiz.\  {\bf 46} (1987) 134].}
   
   
 %\cite{Hawking:1988ae}
\bibitem{Hawking:1988ae}
  S.~W.~Hawking,
  {\em ``Wormholes in Space-Time,''}
  \href{https://journals.aps.org/prd/abstract/10.1103/PhysRevD.37.904}{Phys.\ Rev.\ D {\bf 37} (1988) 904}.
  %doi:10.1103/PhysRevD.37.904
  %%CITATION = doi:10.1103/PhysRevD.37.904;%%
  %412 citations counted in INSPIRE as of 20 Nov 2018  
   
   
   
   %\cite{Giddings:1988cx}
\bibitem{Giddings:1988cx}
  S.~B.~Giddings and A.~Strominger,
  {\em ``Loss of Incoherence and Determination of Coupling Constants in Quantum Gravity,''}
  \href{https://www.sciencedirect.com/science/article/pii/0550321388901095}{Nucl.\ Phys.\ B {\bf 307} (1988) 854.}
   %doi:10.1016/0550-3213(88)90109-5
   
   
   
%\cite{Giddings:1988wv}
\bibitem{Giddings:1988wv}
S.~B.~Giddings and A.~Strominger,
{\em ``Baby Universes, Third Quantization and the Cosmological Constant,''}
\hrj{doi:10.1016/0550-3213(89)90353-2}{Nucl. Phys. B \textbf{321} (1989), 481-508}

%\cite{Krauss:1988zc}
\bibitem{Krauss:1988zc}
L.~M.~Krauss and F.~Wilczek,
{\em ``Discrete Gauge Symmetry in Continuum Theories,''}
\hrj{doi:10.1103/PhysRevLett.62.1221}{Phys. Rev. Lett. \textbf{62} (1989), 1221}


\bibitem{Kallosh:1995hi}
R.~Kallosh, A.~D.~Linde, D.~A.~Linde and L.~Susskind,
{\em ``Gravity and global symmetries,''}
\hrj{doi:10.1103/PhysRevD.52.912}{Phys. Rev. D \textbf{52} (1995), 912-935}
\hri{hep-th/9502069 }{[hep-th]}.

\bibitem{Banks:2010zn}
T.~Banks and N.~Seiberg,
{\em ``Symmetries and Strings in Field Theory and Gravity,''}
\hrj{doi:10.1103/PhysRevD.83.084019}{Phys. Rev. D \textbf{83} (2011), 084019}
\hri{1011.5120}{ [hep-th]}.


%\cite{Maldacena:1997re}
\bibitem{Maldacena:1997re}
J.~M.~Maldacena,
{\em ``The Large N limit of superconformal field theories and supergravity,''}
\hrj{doi:10.1023/A:1026654312961}{Adv. Theor. Math. Phys. \textbf{2} (1998), 231-252}
\hri{hep-th/9711200}{ [hep-th]}.
%17409 citations counted in INSPIRE as of 06 Mar 2022



%\cite{Witten:1998qj}
\bibitem{Witten:1998qj}
  E.~Witten,
  {\em ``Anti-de Sitter space and holography,''}
 \hrj{ doi:10.4310/ATMP.1998.v2.n2.a2}{Adv.\ Theor.\ Math.\ Phys.\  {\bf 2} (1998) 253}
 \hri{hep-th/9802150}{ [hep-th]}.


%\cite{Rey:1998yx}
\bibitem{Rey:1998yx}
S.~J.~Rey,
{\em ``Holographic principle and topology change in string theory,''}
\hrj{10.1088/0264-9381/16/7/102}{Class. Quant. Grav. \textbf{16}, L37-L43 (1999)}
\hri{hep-th/9807241}{ [hep-th]}.
%19 citations counted in INSPIRE as of 23 Oct 2021





%\cite{ArkaniHamed:2007js}
\bibitem{ArkaniHamed:2007js}
  N.~Arkani-Hamed, J.~Orgera and J.~Polchinski,
  {\em ``Euclidean wormholes in string theory,''}
  \href{ https://doi.org/10.1088/1126-6708/2007/12/018}{JHEP {\bf 0712} (2007) 018}
  \hri{0705.2768}{ [hep-th]}.
  %%CITATION = doi:10.1088/1126-6708/2007/12/018;%%
  %60 citations counted in INSPIRE as of 18 Dec 2018
  
 %\cite{Hebecker:2018ofv}
\bibitem{Hebecker:2018ofv}
A.~Hebecker, T.~Mikhail and P.~Soler,
{\em ``Euclidean wormholes, baby universes, and their impact on particle physics and cosmology,''}
\hrj{doi:10.3389/fspas.2018.00035}{Front. Astron. Space Sci. \textbf{5} (2018), 35}
\hri{1807.00824}{ [hep-th]}.
%63 citations counted in INSPIRE as of 28 Mar 2022 


%\cite{Maldacena:2004rf}
\bibitem{Maldacena:2004rf}
  J.~M.~Maldacena and L.~Maoz,
  {\em ``Wormholes in AdS,''}
  \href{https://doi.org/10.1088/1126-6708/2004/02/053}{JHEP {\bf 0402} (2004) 053}
  \hri{hep-th/0401024}.
  
  
  %\cite{Hertog:2018kbz}
\bibitem{Hertog:2018kbz}
T.~Hertog, B.~Truijen and T.~Van Riet,
{\em ``Euclidean axion wormholes have multiple negative modes,''}
\hrj{doi:10.1103/PhysRevLett.123.081302}{Phys. Rev. Lett. \textbf{123} (2019) no.8, 081302}
\hri{1811.12690}{ [hep-th]}.
%20 citations counted in INSPIRE as of 07 Mar 2022


%\cite{Marolf:2020xie}
\bibitem{Marolf:2020xie}
D.~Marolf and H.~Maxfield,
{\em ``Transcending the ensemble: baby universes, spacetime wormholes, and the order and disorder of black hole information,''}
\hrj{doi:10.1007/JHEP08(2020)044}{JHEP \textbf{08} (2020), 044}
\hri{2002.08950}{ [hep-th]}.
%170 citations counted in INSPIRE as of 04 Mar 2022

%\cite{Blommaert:2022ucs}
\bibitem{Blommaert:2022ucs}
A.~Blommaert, L.~V.~Iliesiu and J.~Kruthoff,
{\em ``Alpha states demystified: Towards microscopic models of AdS$_2$ holography,''}
\hri{2203.07384}{ [hep-th]}.
%2 citations counted in INSPIRE as of 27 Mar 2022


%\cite{Johnson:2022wsr}
\bibitem{Johnson:2022wsr}
C.~V.~Johnson,
{\em ``The Microstate Physics of JT Gravity and Supergravity,''}
\hri{2201.11942}{ [hep-th]}.
%4 citations counted in INSPIRE as of 02 Apr 2022
  
 
  
  %\cite{Betzios:2020nry}
\bibitem{Betzios:2020nry}
P.~Betzios and O.~Papadoulaki,
{\em ``Liouville theory and Matrix models: A Wheeler DeWitt perspective,''}
\hrj{10.1007/JHEP09(2020)125}{JHEP \textbf{09} (2020), 125}
\hri{2004.00002}{ [hep-th]}.
%14 citations counted in INSPIRE as of 10 Feb 2021


%\cite{McNamara:2020uza}
\bibitem{McNamara:2020uza}
J.~McNamara and C.~Vafa,
{\em ``Baby Universes, Holography, and the Swampland,''}
\hri{2004.06738}{ [hep-th]}.
%65 citations counted in INSPIRE as of 04 Mar 2022

  



 \bibitem{worm}
P.~Betzios, E.~Kiritsis and O.~Papadoulaki,
{\em ``Euclidean Wormholes and Holography,''}
\hrj{10.1007/JHEP06(2019)042}{JHEP \textbf{06} (2019), 042};
\hri{1903.05658}{ [hep-th]}.
%12 citations counted in INSPIRE as of 31 Oct 2020


%\cite{Betzios:2021fnm}
\bibitem{Betzios:2021fnm}
P.~Betzios, E.~Kiritsis and O.~Papadoulaki,
{\em ``Interacting systems and wormholes,''}
\hrj{doi:10.1007/JHEP02(2022)126}{JHEP \textbf{02} (2022), 126}
\hri{2110.14655}{ [hep-th]}.
%2 citations counted in INSPIRE as of 04 Mar 2022



%\cite{VanRaamsdonk:2020tlr}
\bibitem{VanRaamsdonk:2020tlr}
M.~Van Raamsdonk,
{\em ``Comments on wormholes, ensembles, and cosmology,''}
\hri{2008.02259}{ [hep-th]} ; \\
%13 citations counted in INSPIRE as of 17 Jan 2021
{\em ``Cosmology from confinement?,''} \\
\hrj{doi:10.1007/JHEP03(2022)039}{JHEP \textbf{03} (2022), 039}
\hri{2102.05057}{ [hep-th]}.
%15 citations counted in INSPIRE as of 28 Mar 2022

%\cite{Kundu:2021nwp}
\bibitem{Kundu:2021nwp}
A.~Kundu,
{\em ``Wormholes \& Holography: An Introduction,''}
[arXiv:2110.14958 [hep-th]].


\bibitem{Saad:2019lba}
P.~Saad, S.~H.~Shenker and D.~Stanford,
{\em ``JT gravity as a matrix integral,''}
\hri{1903.11115}{ [hep-th]}.



%\cite{Seiberg:1990eb}
\bibitem{Seiberg:1990eb}
N.~Seiberg,
{\em ``Notes on quantum Liouville theory and quantum gravity,''}
\hrj{doi:10.1143/PTPS.102.319}{Prog. Theor. Phys. Suppl. \textbf{102} (1990), 319-349}
%422 citations counted in INSPIRE as of 05 Mar 2022



%\cite{tHooft:2014swy}
\bibitem{tHooft:2014swy}
G.~'t Hooft,
{\em ``Local Conformal Symmetry: the Missing Symmetry Component for Space and Time,''}
\hri{1410.6675}{ [gr-qc]}.



\bibitem{Betzios:2017krj}
P.~Betzios, N.~Gaddam and O.~Papadoulaki,
{\em ``Antipodal correlation on the meron wormhole and a bang-crunch universe,''}
\hrj{doi:10.1103/PhysRevD.97.126006}{Phys. Rev. D \textbf{97} (2018) no.12, 126006}
\hri{1711.03469}{ [hep-th]}.


%\cite{Mtheory}
\bibitem{Mtheory}
L.~Thorlacius and T.~Jonsson, eds., 
{\em ``M-theory and Quantum Geometry,"} 
\href{https://link.springer.com/book/10.1007/978-94-011-4303-5}{Springer Science and Business Media (Vol. 556) 2000.}



\bibitem{Tong:2014era}
D.~Tong and C.~Turner,
{\em ``Quantum dynamics of supergravity on R$^3 \times$ S$^1$,''}
\hrj{doi:10.1007/JHEP12(2014)142}{JHEP \textbf{12} (2014), 142}
\hri{1408.3418}{ [hep-th]}.

%\cite{Marolf:2021kjc}
\bibitem{Marolf:2021kjc}
D.~Marolf and J.~E.~Santos,
{\em ``AdS Euclidean wormholes,''}
\hrj{doi:10.1088/1361-6382/ac2cb7}{Class. Quant. Grav. \textbf{38} (2021) no.22, 224002}
\hri{2101.08875}{ [hep-th]}.
%24 citations counted in INSPIRE as of 28 Mar 2022

\bibitem{Loges:2022nuw}
G.~J.~Loges, G.~Shiu and N.~Sudhir,
{\em ``Complex Saddles and Euclidean Wormholes in the Lorentzian Path Integral,''}
\hri{2203.01956}{ [hep-th]}.


%\cite{Lunin:2006xr}
\bibitem{Lunin:2006xr}
O.~Lunin,
{\em ``On gravitational description of Wilson lines,''}
\hrj{doi:10.1088/1126-6708/2006/06/026}{JHEP \textbf{06} (2006), 026}
\hri{hep-th/0604133}{ [hep-th]}.
%183 citations counted in INSPIRE as of 28 Mar 2022


%\cite{Iqbal:2003ds}
\bibitem{Iqbal:2003ds}
A.~Iqbal, N.~Nekrasov, A.~Okounkov and C.~Vafa,
{\em ``Quantum foam and topological strings,''}
\hrj{doi:10.1088/1126-6708/2008/04/011}{JHEP \textbf{04} (2008), 011}
\hri{hep-th/0312022}{ [hep-th]}.
%223 citations counted in INSPIRE as of 27 Mar 2022

%\cite{Dijkgraaf:2005bp}
\bibitem{Dijkgraaf:2005bp}
R.~Dijkgraaf, R.~Gopakumar, H.~Ooguri and C.~Vafa,
{\em ``Baby universes in string theory,''}
\hrj{doi:10.1103/PhysRevD.73.066002}{Phys. Rev. D \textbf{73} (2006), 066002}
\hri{hep-th/0504221}{ [hep-th]}.
%94 citations counted in INSPIRE as of 04 Mar 2022


%\cite{Lin:2004nb}
\bibitem{Lin:2004nb}
H.~Lin, O.~Lunin and J.~M.~Maldacena,
{\em ``Bubbling AdS space and 1/2 BPS geometries,''}
\hrj{doi:10.1088/1126-6708/2004/10/025}{JHEP \textbf{10} (2004), 025}
\hri{hep-th/0409174}{[hep-th]}.
%817 citations counted in INSPIRE as of 27 Mar 2022



%\cite{Betzios:2017yms}
\bibitem{Betzios:2017yms}
P.~Betzios and O.~Papadoulaki,
{\em ``FZZT branes and non-singlets of matrix quantum mechanics,''}
\hrj{10.1007/JHEP07(2020)157}{JHEP \textbf{07} (2020), 157}
\hri{1711.04369}{ [hep-th]}.
%6 citations counted in INSPIRE as of 10 Feb 2021


%\cite{Douglas:1993iia}
\bibitem{Douglas:1993iia}
M.~R.~Douglas and V.~A.~Kazakov,
{\em ``Large N phase transition in continuum QCD in two-dimensions,''}
\hrj{doi:10.1016/0370-2693(93)90806-S}{Phys. Lett. B \textbf{319} (1993), 219-230}
\hri{hep-th/9305047}{ [hep-th]}.
%215 citations counted in INSPIRE as of 16 Mar 2022







%\cite{McNamara:2019rup}
\bibitem{McNamara:2019rup}
J.~McNamara and C.~Vafa,
{\em ``Cobordism Classes and the Swampland,''}
\hri{1909.10355}{ [hep-th]}.
%59 citations counted in INSPIRE as of 04 Mar 2022






%\cite{Casini:2021zgr}
\bibitem{Casini:2021zgr}
H.~Casini and J.~M.~Magan,
{\em ``On completeness and generalized symmetries in quantum field theory,''}
\hrj{doi:10.1142/S0217732321300251}{Mod. Phys. Lett. A \textbf{36} (2021) no.36, 2130025}
\hri{2110.11358}{ [hep-th]}.
%6 citations counted in INSPIRE as of 04 Mar 2022


%\cite{McNamara:2021cuo}
\bibitem{McNamara:2021cuo}
J.~McNamara,
{\em ``Gravitational Solitons and Completeness,''}
\hri{2108.02228}{ [hep-th]}.
%4 citations counted in INSPIRE as of 04 Mar 2022





\bibitem{Cecotti:2010bp}
S.~Cecotti, C.~Cordova, J.~J.~Heckman and C.~Vafa,
{\em ``T-Branes and Monodromy,''}
\hrj{doi:10.1007/JHEP07(2011)030}{JHEP \textbf{07} (2011), 030}
\hri{1010.5780}{ [hep-th]}.



\bibitem{Heckman:2021vzx}
J.~J.~Heckman, A.~P.~Turner and X.~Yu,
{\em ``Disorder Averaging and its UV (Dis)Contents,''}
\hri{2111.06404}{ [hep-th]}.








\bibitem{Gaiotto:2008ak}
D.~Gaiotto and E.~Witten,
{\em ``S-Duality of Boundary Conditions In N=4 Super Yang-Mills Theory,''}
\hrj{doi:10.4310/ATMP.2009.v13.n3.a5}{Adv. Theor. Math. Phys. \textbf{13} (2009) no.3, 721-896}
\hri{0807.3720}{ [hep-th]}.






%\cite{Fischler:1988ia}
\bibitem{Fischler:1988ia}
W.~Fischler and L.~Susskind,
{\em ``A WORMHOLE CATASTROPHE,''}
\hrj{doi:10.1016/0370-2693(89)91514-1}{Phys. Lett. B \textbf{217} (1989), 48-54}
%123 citations counted in INSPIRE as of 16 Mar 2022





%\cite{Maldacena:2001kr}
\bibitem{Maldacena:2001kr}
J.~M.~Maldacena,
{\em ``Eternal black holes in anti-de Sitter,''}
\hrj{doi:10.1088/1126-6708/2003/04/021}{JHEP \textbf{04} (2003), 021}
\hri{hep-th/0106112}{ [hep-th]}.
%1140 citations counted in INSPIRE as of 15 Mar 2022


%\cite{VanRaamsdonk:2010pw}
\bibitem{VanRaamsdonk:2010pw}
M.~Van Raamsdonk,
{\em ``Building up spacetime with quantum entanglement,''}
\hrj{doi:10.1142/S0218271810018529}{Gen. Rel. Grav. \textbf{42} (2010), 2323-2329}
\hri{1005.3035}{ [hep-th]}.







\end{thebibliography}
\end{document}